\documentclass[aps,superscriptaddress,showpacs,letterpaper,twocolumn,reprint,nofootinbib,longbibliography,pra]{revtex4-2}
\usepackage{graphicx} 
\usepackage{orcidlink}
\usepackage{amsmath} 
\usepackage{amssymb}   
\usepackage{color}  
\usepackage{enumitem}
\usepackage{listings} 
\newcommand{\rvec}{\mathbf{r}}
\newcommand{\Evec}{\mathbf{E}}

\newcommand{\ketbra}[2]{|#1 \rangle \langle #2 |}

\newcommand{\Heff}{H_{\mathrm{eff}}}

\newcommand{\ket}[1]{|#1\rangle}

\newcommand \be{\begin{equation}}
\newcommand \ee{\end{equation}}
\newcommand \bea{\begin{eqnarray}}
\newcommand \eea{\end{eqnarray}}

\begin{document}

\title{Three-dimensional three-photon Stark spectroscopy of a single Rb Rydberg  atom in an ultrahigh-vacuum glass cell with eight electrodes}

\author{S.~A.~Spirin~\orcidlink{0009-0003-5224-8454}}
\affiliation{Rzhanov Institute of Semiconductor Physics SB RAS, 630090 Novosibirsk, Russia}
\affiliation{Novosibirsk State University, 630090 Novosibirsk, Russia}

\author{E.~A.~Yakshina~\orcidlink{0009-0000-7621-2454}}
\affiliation{Rzhanov Institute of Semiconductor Physics SB RAS, 630090 Novosibirsk, Russia}
\affiliation{Novosibirsk State University, 630090 Novosibirsk, Russia}
\affiliation{Institute of Laser Physics SB RAS, 630090 Novosibirsk, Russia}

\author{A.~K.~Sologub~\orcidlink{0009-0007-2725-9687}}
\affiliation{Rzhanov Institute of Semiconductor Physics SB RAS, 630090 Novosibirsk, Russia}
\affiliation{Novosibirsk State University, 630090 Novosibirsk, Russia}

\author{G.~Suliman~\orcidlink{0009-0000-3769-4235}}
\affiliation{Rzhanov Institute of Semiconductor Physics SB RAS, 630090 Novosibirsk, Russia}
\affiliation{Novosibirsk State University, 630090 Novosibirsk, Russia}

\author{A.~M.~Minnigaliev~\orcidlink{0009-0001-7871-9408}}
\affiliation{Rzhanov Institute of Semiconductor Physics SB RAS, 630090 Novosibirsk, Russia}

\author{V.~V.~Gromyko~\orcidlink{0009-0004-5026-3535}}
\affiliation{Rzhanov Institute of Semiconductor Physics SB RAS, 630090 Novosibirsk, Russia}
\affiliation{Institute of Laser Physics SB RAS, 630090 Novosibirsk, Russia}

\author{P.~I.~Betleni~\orcidlink{0009-0006-2964-6229}}
\affiliation{Rzhanov Institute of Semiconductor Physics SB RAS, 630090 Novosibirsk, Russia}
\affiliation{Novosibirsk State University, 630090 Novosibirsk, Russia}

\author{D.~B.~Tretyakov~\orcidlink{0000-0002-3708-6253}}
\affiliation{Rzhanov Institute of Semiconductor Physics SB RAS, 630090 Novosibirsk, Russia}

\author{V.~M.~Entin~\orcidlink{0000-0001-5436-2849}}
\affiliation{Rzhanov Institute of Semiconductor Physics SB RAS, 630090 Novosibirsk, Russia}

\author{N.~N.~Bezuglov~\orcidlink{0000-0003-0191-988X}}
\affiliation{Saint-Petersburg State University, 199034 St.Petersburg, Russia}
\affiliation{Rzhanov Institute of Semiconductor Physics SB RAS, 630090 Novosibirsk, Russia}

\author{I.~I.~Beterov~\orcidlink{0000-0002-6596-6741}}
\affiliation{Rzhanov Institute of Semiconductor Physics SB RAS, 630090 Novosibirsk, Russia}
\affiliation{Novosibirsk State University, 630090 Novosibirsk, Russia}
\affiliation{Institute of Laser Physics SB RAS, 630090 Novosibirsk, Russia}
\affiliation{Novosibirsk State Technical University, 630073 Novosibirsk, Russia}

\author{I.~I.~Ryabtsev~\orcidlink{0000-0002-5410-2155}}
\affiliation{Rzhanov Institute of Semiconductor Physics SB RAS, 630090 Novosibirsk, Russia}
\affiliation{Novosibirsk State University, 630090 Novosibirsk, Russia}

\date{\today}

\begin{abstract} 
Quantum computing and quantum simulation with ultracold neutral atoms require Rydberg excitation of individual atoms in atomic arrays. Rydberg states are extremely sensitive to external electric fields, therefore precise three-dimensional control of the electric field  is essential. We performed a spectroscopic study of three-photon Rydberg excitation of a single Rb atom in an optical dipole trap in the presence of an external DC electric field. The field was generated by eight electrodes deposited on the inner surfaces of an ultrahigh-vacuum glass cell. The used three-photon scheme of laser excitation of Rydberg \textit{nP} states allows the Stark shift and the splitting of the resonances to be observed simultaneously, which simplifies calibration of the electric field. In addition, in the commonly used two-photon Rydberg excitation schemes, the light shifts can complicate accurate determination of the DC Stark shift, particularly when the external electric field is scanned across different spatial directions, and different Stark components are excited. Our analytical expressions show that these shifts are suppressed in the three-photon excitation scheme. We demonstrated the ability to independently tune the electric field along all three spatial directions and to compensate for stray electric fields. The measured three-photon spectra exhibit Stark shifts and splittings of the three-photon resonance that are in good agreement with theoretical calculations. These results are also of interest for Rydberg electrometry.

\end{abstract}

\maketitle

\section{Introduction}

Large-scale arrays of ultracold neutral atoms currently represent one of the most promising platforms for universal quantum computers and quantum simulators. The ability to trap thousands of atoms in arrays of optical dipole traps has been demonstrated in recent years~\cite{Chiu2025, Manetsch2024, Gyger2024, Pichard2024, Norcia2024}. Entanglement between neutral atoms is typically generated using long-range Rydberg interactions~\cite{Jaksch2000}. When excited to Rydberg states, atoms interact strongly at distances of a few micrometers, matching the typical interatomic spacing in such experiments. Due to their large polarizabilities, Rydberg states are highly sensitive to external electric fields~\cite{Gallagher1994}. This sensitivity enables numerous applications such as Rydberg electrometry~\cite{Osterwalder1999,Thiele2015,Grimmel2015,Grimmel2017,Kitson2026,Zhang2024, Meyer2020, Degen2017} and tunable interatomic interactions~\cite{Ryabtsev2010, Nipper2012, Tretyakov2014, Jiao2022}. However, stray electric fields can shift and broaden Rydberg resonances, thereby degrading the coherence of atom-light interactions. Precise three-dimensional control of the electric field is therefore essential for quantum computing and quantum simulation experiments with large-scale atomic ensembles. It is also of great importance for Rydberg electrometry.

Stark spectroscopy of alkali-metal Rydberg states is a very established research area that has many applications. The scalar polarizabilities of the Rb Rydberg states were  measured for the first time using a gas discharge lamp combined with a dye laser~\cite{Fredriksson1977}, and later using  Doppler-free two-photon laser spectroscopy in a vapor cell~\cite{Sullivan1985}. The first calculations of the Stark maps of Rydberg energy levels and a comparison with the experiment in lithium and cesium atomic beams using selective field ionization detector~\cite{Ducas1975, Gallagher1977} for detection of Rydberg atoms were published in a classical work by Zimmerman et al.~\cite{Zimmerman1979}. The three pairs of planar electrodes were used to compensate for the for the stray electric field in an atom beam setup~\cite{Frey1993}. The atomic beam technique was used both in earlier~\cite{Beterov1998,Ryabtsev2000,Ryabtsev2003} and later experiments~\cite{Hiller2014, Zhang2015,Yoshida2017}.  Rydberg atoms nowadays  experience increased interest as electric field sensors~\cite{Osterwalder1999,Thiele2015,Grimmel2015,Grimmel2017,Kitson2026,Zhang2024, Meyer2020}. Recently, the atomic beam and selective field ionization technique were used for design of a low-frequency electric field sensor with Rb Rydberg atoms~\cite{Glick2026}. 

With the development of methods of laser cooling and trapping Stark spectroscopy of Rydberg excitation of ultracold atoms became possible~\cite{Grabowski2006,Vogt2007,Viteau2011}.  In particular, the splitting of Stark components of Rb 41\textit{D} state with different values of $m_j$ was studied in the magneto-optical trap (MOT)~\cite{Grabowski2006}. The quenching of Rydberg population due to Rydberg blockade in the external electric field was experimentally observed in a MOT~\cite{Vogt2007}. Later, the Stark energy shifts allowed for investigation of F\"{o}rster resonances with single Rydberg atoms in a MOT using selective field ionization~\cite{Ryabtsev2010, Nipper2012, Ravets2014,Tretyakov2014, Jiao2022}, and later in the single-atom experiment in optical dipole traps with optical detection of Rydberg atoms~\cite{Anand2024}.  Careful analysis and compensation of stray electric fields in a MOT using a system of eight segmented ring electrodes was reported in Ref.~\cite{Sassmannshausen2013}. The experiment was based on microwave spectroscopy of transitions between \textsuperscript{133}Cs Rydberg states and selective field ionization. A single-photon laser excitation scheme  was used for Stark spectroscopy of Cs atoms~\cite{Bai2020}. Two-photon laser excitation in the DC electric field of Rydberg \textit{nS} and \textit{nD} states of magnetically trapped Rb atoms near the atom chip surface  was studied~\cite{Cisternas2017}.   Recently, interest in the behavior of ultracold Rydberg atoms in strong electric fields has increased~\cite{Stecker2020}.

Compared to selective field ionization, a much simpler, but slower experimental approach for observation of Stark shifts in ultracold Rydberg atoms is the trap-loss spectroscopy~\cite{Zelener2015,Halter2023}. If the lasers are tuned on transitions to Rydberg states, the atoms are removed from the MOT, since they stop interacting with the cooling radiation. To determine the losses of atoms, the fluorescence images of the MOT are taken. This method was used to measure the polarizabilities of ytterbium Rydberg states~\cite{Halter2023}. A similar approach allowed to eliminate the stray electric field in all three dimensions~\cite{Panja2024}. 

The single-atom losses are also being used for the detection of Rydberg excitation in experiments with single ultracold atoms in optical dipole traps~\cite{Johnson2008}. When an atom is excited into the Rydberg state, the intensive radiation of the dipole trap repels it out of the trap~\cite{Saffman2010}. In a single-atom experiment~\cite{Ravets2014} the electric field generated by eight segmented ring electrodes was used to tune the F\"{o}rster resonances between two 59\textit{D}\textsubscript{3/2} Rb Rydberg atoms. That experiment was performed in a metal vacuum chamber using a two-photon Rydberg excitation scheme. Later, a similar approach was used to study the heteronuclear Rb-Cs interaction~\cite{Anand2024}. 

A three-photon scheme of laser excitation has several advantages. First, laser excitation of Rydberg $nP$ states allows for simultaneous observation of shifts and splittings of the resonances, compared to the experiments with $nS$ states. Moreover, in most cases $nP$ Rydberg states  possess larger static polarizabilities compared to $nS$ states. The two-photon excitation of Rydberg $nD$ states may have similar advantages over experiments with $nS$ states, but such experiments are complicated by the richer Stark structure and larger light shifts of the resonances, that are observed in two-photon excitation schemes. 

When the electric field is scanned across different spatial directions, laser excitation of different Stark components of the Rydberg state results in variation of the Rabi frequency of the second step of laser excitation, meaning effectively changing the polarization of laser radiation with respect to the quantization axis. Thus, it changes the light shifts experienced by the atoms. This undesirable effect is suppressed in a three-photon laser excitation scheme with strong intermediate coupling, proposed by us recently~\cite{Bezuglov2025}. 

Many modern experiments in laser cooling, quantum information processing, and metrology, including those with large-scale atomic arrays, are performed in ultrahigh-vacuum (UHV) glass cells. Compared to metal vacuum chambers, these cells provide significantly better optical access, a more compact design, and the ability to rapidly switch off the magnetic field gradients required for magneto-optical traps. The cells are typically of either simple rectangular or octagonal multi-facet designs. The latter geometry offers superior optical access and allows laser beams, imaging optics, and magnetic coils to be placed closer to the atoms.

Achieving three-dimensional electric-field control inside such cells is challenging. High-numerical-aperture lenses for the trapping beams, which must be located outside the cell, together with the laser cooling beams, demand unobstructed optical access. Because the glass walls become polarized under an applied voltage, electrodes placed outside the cell are inefficient at generating well-controlled DC fields inside. Consequently, electrodes must be placed inside the vacuum cell.

\begin{figure}[!t]
\includegraphics[width=\columnwidth]{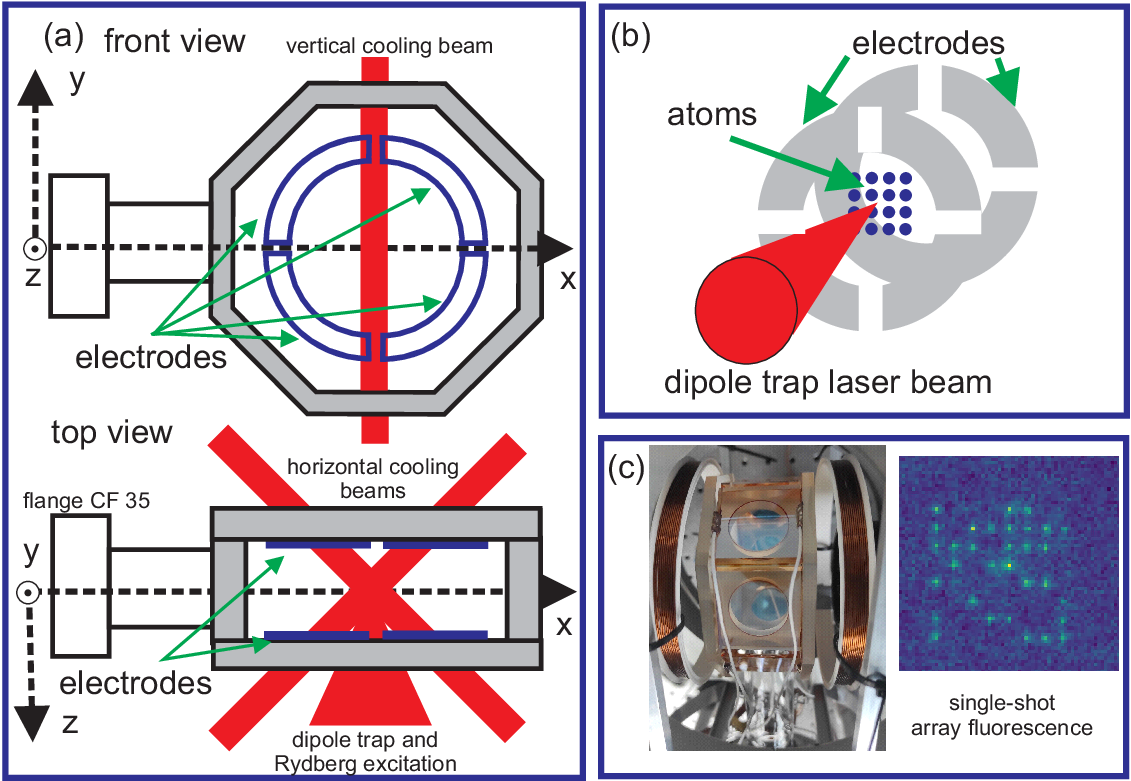}
\caption{ (a) Design of the UHV vacuum glass cell for our experiments with arrays of ultracold neutral atoms. The octagonal multi-facet cell is placed vertically. The large round window on the front side of the cell is used simultaneously for passing the horizontal  MOT cooling beams and the beam of the optical dipole trap. The vertical cooling beam is sent through two facets of the octagon on the opposite sides. The coaxial ring-shape electrodes are deposited on the inner surface of the glass cell. Each ring has four identical segments. (b) Sketch of the atomic array located between the electrodes. (c) Photo of the UHV glass cell in the experimental setup and the single-shot fluorescence image of the single-atom array. The coils for compensation of stray magnetic field are visible on the photo.
}
\label{Setup}
\end{figure}

We therefore commissioned a custom octagonal UHV quartz cell from CAS Cold Atom\textsuperscript{TM}. Its design is shown in Fig.~\ref{Setup}(a). The octagonal cell is placed vertically. The horizontal MOT beams are sent, along with the optical dipole trap beam, through the large front window. Eight thin electrodes were deposited on the inner surface of the cell in the form of two concentric rings. Each ring is divided into four equal sections, and an independent voltage can be applied to each of the eight electrodes.
Similar electrode geometry is used in many experiments with ultracold Rydberg atoms, but mostly in metal vacuum chambers~\cite{Sassmannshausen2013,Ravets2014, Ding2018, Hofmann2014}. The position of the atomic array between the electrodes is illustrated in  Fig.~\ref{Setup}(b). The photo of the cell in the assembled experimental setup and the single-shot fluorescence image of the randomly loaded single-atom atomic array are shown in Fig.~\ref{Setup}(c). To verify the ability for precise individual control of the electric field in any direction, we performed experiments on Stark spectroscopy of three-photon Rydberg excitation of a single Rb atom in the external DC electric field.

The suppression of light shifts of the three-photon Rydberg excitation resonances allowed us to perform a sequence of experiments in the same conditions, and indvidually calibrate the electric field along each coordinate. We found that the stray electric fields in the experiment may reach substantially high values of several hundred millivolts per centimeter. However, they can be suppressed using the electrodes.

The paper is organized as follows.   In Sec.~\ref{sec.Theory} we present the theoretical framework for calculation of the spectra of three-photon Rydberg excitation in the external DC electric field and discuss the three-photon and two-photon light shifts. In Sec.~\ref{sec.Experiment} we describe the scheme of the experimental setup and methods. In Sec.~\ref{sec.Results} we compare recorded spectra with the theory. The results are discussed and summarized in Secs.~\ref{sec.Discussion} and \ref{sec.Conclusion}. The details of derivation of the analytical expressions for three-photon light shifts and calculation of the electric field are listed in the Appendices~\ref{sec.Shifts}, \ref{sec.Field}.

\section{Theory of three-photon Stark spectroscopy}
\label{sec.Theory}

\begin{figure}[!t]
\includegraphics[width= \columnwidth]{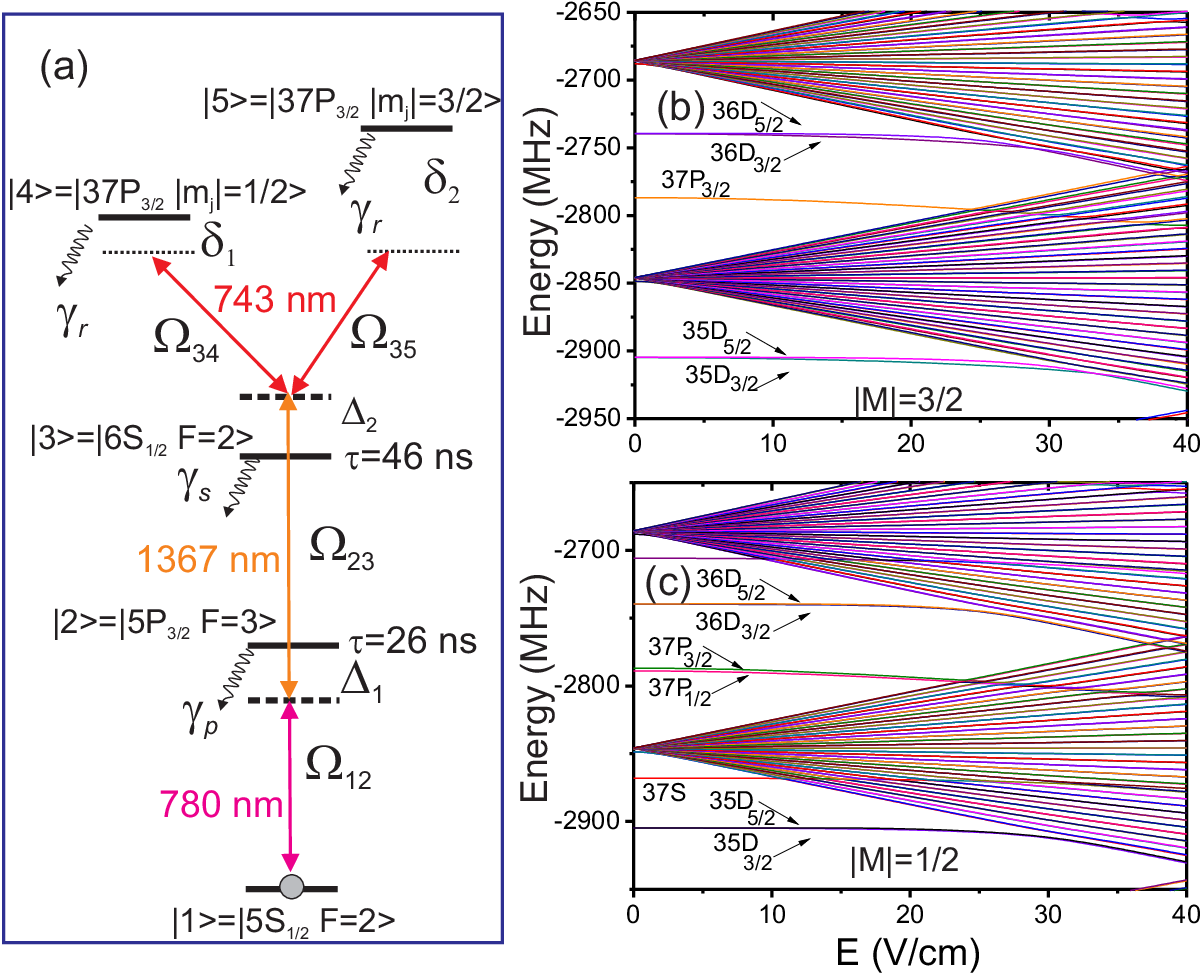}
\caption{(a) Scheme of the three-photon Rydberg excitation of two Stark sublevels of the Rb $37P_{3/2}$ state. (b),(c) Numerically calculated Stark diagrams for  (b) $|m_j|=3/2$ and  (c) $|m_j|=1/2$ indicate a quadratic Stark shift of the Rydberg $37P_{3/2}$ state.
\label{StarkDiagram}
}
\end{figure}

The scheme of the three-photon Rydberg excitation of two Stark sublevels of Rb Rydberg $37P_{3/2}$ state is shown in Fig.~\ref{StarkDiagram}(a). We consider a five-level atom with ground state $\ket{1}=\ket{5S_{1/2}\, F=2}$, intermediate excited states $\ket{2}=\ket{5P_{3/2}\, F=3}$, $\ket{3}=\ket{6S_{1/2}\, F=2}$ and two Stark sublevels of Rydberg state $\ket{4}=\ket{37P_{3/2}\, |m_j|=1/2}$ and $\ket{5}=\ket{37P_{3/2}\, |m_j|=3/2}$.

The lower part of the system therefore forms a ladder $\ket{1} \to \ket{2} \to \ket{3}$,
while the state $\ket{3}$ is coupled to two upper states $\ket{4}$, and $\ket{5}$ depending on the polarization of laser radiation.

The three laser fields interact with an atom. The first laser field with wavelength 780~nm and Rabi frequency $\Omega_{12}$ drives the transition $\ket{1} \to \ket{2}$ with red detuning $\Delta_1=10$~MHz from the resonance, which is kept constant. The second laser field with wavelength~1367 nm and Rabi frequency $\Omega_{12}$ drives the transition $\ket{1} \to \ket{2}$ with blue two-photon detuning $\Delta_2=-200$~MHz. The upper excitation stage couples the intermediate state $\ket{3}$ to the two upper states $\ket{4}$  and $\ket{5}$, with Rabi frequencies $\Omega_{34}$ and $\Omega_{35}$, and variable detunings $\delta_1$ and $\delta_2$, respectively. The latter detunings depend on the Stark shift of each of the Stark components of the $37P_{3/2}$ state in the external DC electric field. The numerically calculated Stark diagrams for Rb Rydberg states using a quasiclassical method~\cite{Kaulakys1995} of calculation of radial matrix elements are shown in Fig.~\ref{StarkDiagram}(b) for $|m_j|=3/2$  and in Fig.~\ref{StarkDiagram}(c) for $|m_j|=1/2$.  The Stark components of the $nP_{3/2}$ Rydberg states in Rb exhibit a quadratic Stark shift $-\frac{1}{2}\alpha E^2$, which is defined by the polarizabilities $\alpha_{P1/2}=38.8$~MHz/(V/cm)\textsuperscript{2} and $\alpha_{P3/2}=32.5$~MHz/(V/cm)\textsuperscript{2}. These values are obtained from quadratic approximation of the numerically calculated Stark diagrams for the smaller values of the electric field below 2~V/cm.

In the rotating-wave approximation, and in the interaction picture defined by the laser frequencies, the Hamiltonian can be written as
  
\begin{eqnarray}
&&H=\hbar\left(\Delta_1 \ketbra{2}{2}+\Delta_2 \ketbra{3}{3}+\delta_1 \ketbra{4}{4}
+\delta_2 \ketbra{5}{5}\right)+  \\
&&+\frac{\hbar}{2}\left[\Omega_{12}\ketbra{1}{2}+\Omega_{23}\ketbra{2}{3}+\Omega_{34}\ketbra{3}{4}+\Omega_{35}\ketbra{3}{5}+\mathrm{H.c.}\right]\nonumber
\end{eqnarray}

The diagonal terms represent the detunings in the rotating frame, while the off-diagonal terms describe the coherent laser couplings.

The density operator \(\rho\) evolves according to the Lindblad master equation
\be
\frac{d\rho}{dt}
=
-i[H,\rho]
+\sum_k
\left(
L_k \rho L_k^\dagger
-\frac{1}{2}\{L_k^\dagger L_k,\rho\}
\right),
\ee
where the collapse operators $L_k$ account for spontaneous decay and laser-induced dephasing.

Spontaneous emission is included through the finite lifetimes of the excited states which are 26~ns for $5P_{3/2}$ state, 46~ns for $6S_{1/2}$ state~\cite{Vsibalic2017ARC} and 43~$\mu$s for $37P_{3/2}$ state~\cite{Vsibalic2017ARC,Beterov2009}. 
The atom is assumed to be initially prepared in the ground state,
$\rho(0)=|1\rangle\langle 1|$. It interacts with the laser fields for a fixed time $T=4\,\mu$s, after which only the final state $\rho(T)$ is used. The level populations at the end of the interaction are
$P_i(T)=\langle i|\rho(T)|i\rangle,\qquad i=1,\dots,5$.
The experimentally relevant signal is taken to be the total population in the two upper Rydberg states, $P_{45}(T)=P_4(T)+P_5(T)$.

An important advantage of the three-photon laser excitation scheme considered in this work is the ability to completely suppress the light shifts that commonly arise in two-photon excitation schemes with imbalanced Rabi frequencies~\cite{Bezuglov2025}. For the two-photon excitation scheme $\ket{1} \to \ket{2} \to \ket{3} $ with coupling Rabi frequencies  $\Omega_{1}$ and $\Omega_{2}$ and detuning  $\Delta$ from the intermediate state the two-photon resonance is  AC Stark shifted by a value $\delta_2^{\mathrm{res}}=(\Omega_{2}^2-\Omega_{1}^2)/(4\Delta)$. 

\begin{figure}[!t]
\includegraphics[width= \columnwidth]{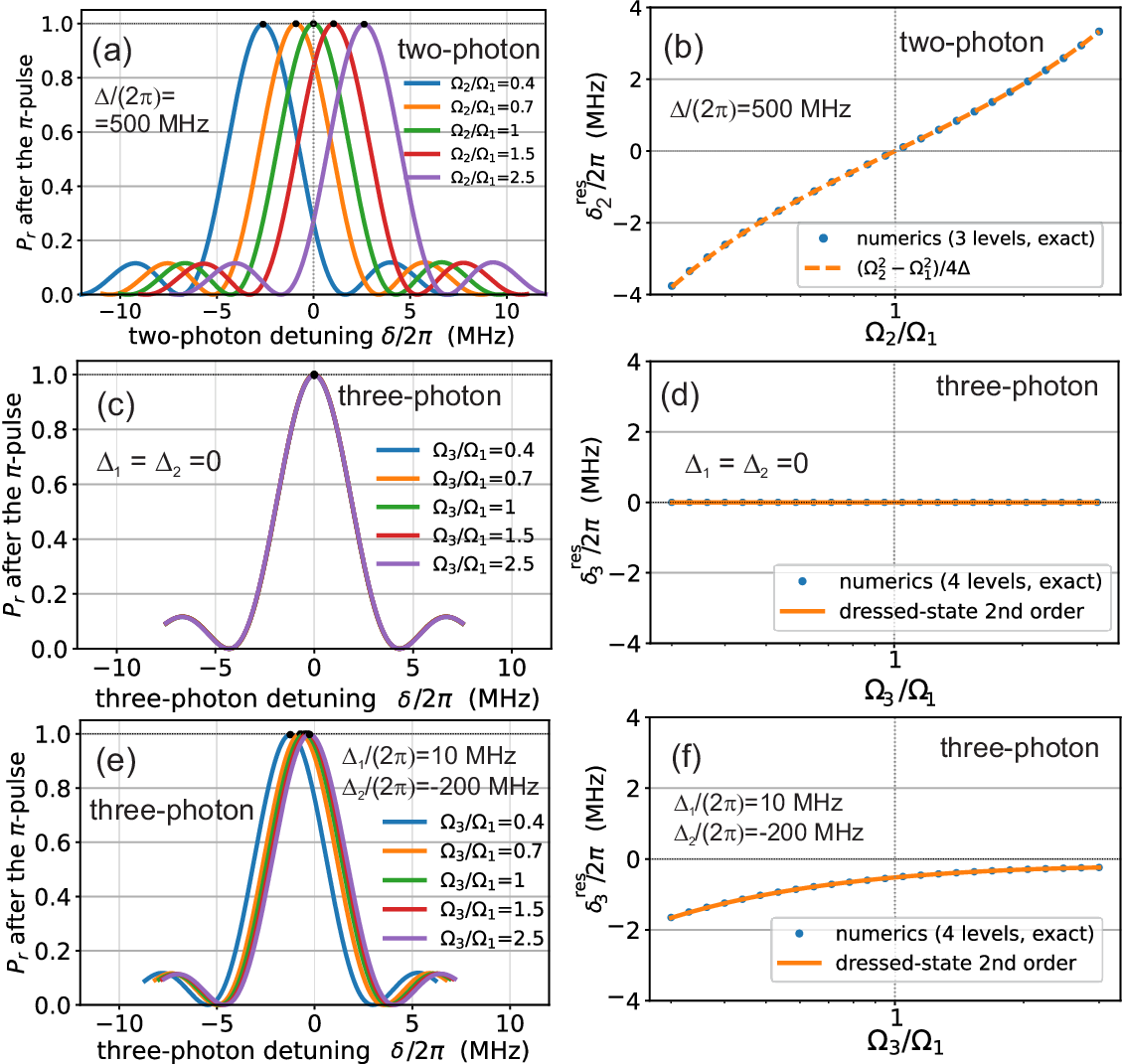}
\caption{Comparison of the light shifts for two-photon and three-photon Rydberg excitation schemes for a laser $\pi$-pulse  with 200~ns duration. (a),(b) Numerically calculated single-atom excitation probabilities $P_1$ and light shifts $\delta_2^{\mathrm{res}}$ of two-photon Rydberg excitation with 500~MHz detuning from the intermediate state for different ratios of Rabi frequencies $\Omega_2/\Omega_1$. (c),(d) Numerically calculated single-atom excitation probabilities $P_1$ and light shifts $\delta_3^{\mathrm{res}}$ of three-photon Rydberg excitation with coupling $\Omega_2/(2\pi)=1000$~MHz  for different ratios of Rabi frequencies $\Omega_3/\Omega_1$ at exact three-photon resonance with $\Delta_1=\Delta_2=0$.  (e),(f) Numerically calculated  single-atom excitation probabilities $P_1$ and light shifts $\delta_3^{\mathrm{res}}$ of three-photon Rydberg excitation with  coupling $\Omega_2/(2\pi)=1000$~MHz  for different ratios of Rabi frequencies $\Omega_3/\Omega_1$ at $\Delta_1/(2\pi)=10$~MHz  and $\Delta_2/(2\pi)=-200$~MHz.}
\label{TwoVsThree}
\end{figure}

Figure~\ref{TwoVsThree}(a) illustrates the shift of the theoretically calculated spectra of two-photon Rydberg excitation by a laser $\pi$-pulse with 200~ns duration for different ratios $\Omega_{2}/\Omega_{1}$. Here we take a 500~MHz detuning from the intermediate excited state and vary the ratio of Rabi frequencies starting with $\Omega_{1}/(2\pi)=\Omega_{2}/(2\pi)=50$~MHz, while keeping the same duration of the laser $\pi$-pulse. We neglected the finite lifetimes of the Rydberg and intermediate excited states. The comparison of the numerically calculated dependence of the shift of the two-photon resonance with the analytical expression is shown in Fig.~\ref{TwoVsThree}(b). 

For three-photon Rydberg excitation,  in the regime of strong coupling between the intermediate levels $\Omega_2 \gg  \Omega_1, \Omega_3$, single-photon detuning $\Delta_1$ and two-photon detuning $\Delta_2$ from the intermediate excited states, the  three-photon resonance shift is given by the expression

\be
\delta_3^{\mathrm{res}}=\frac{\Omega_3^{2}\Delta_1-\Omega_1^{2}\Delta_2}{4\Delta_1\Delta_2-\Omega_2^{2}}.
\label{eq:shifts}
\ee 
We derive it in Appendix~\ref{sec.Shifts}. The same values of the Rabi frequencies $\Omega_{1}/(2\pi)=\Omega_{3}/(2\pi)=50$~MHz will give   200~ns  $\pi$-pulse duration, which makes the conditions similar to the previously considered two-photon case.

Figures~\ref{TwoVsThree}(c),(d) show that for $\Omega_2/(2\pi)=1000$~MHz and zero detunings from the intermediate excited state the light shifts are completely suppressed.   Due to the AC Stark splitting of the intermediate excited states, even at zero detunings from the intermediate excited states it is possible to achieve coherent three-photon Rydberg excitation without populating the intermediate excited states~\cite{Bezuglov2025}, and observe coherent Rabi flopping between ground and Rydberg levels~\cite{Beterov2024}.

However, in our experiment, due to the technical limitations of the laser locking  we used non-zero intermediate detunings. Figures~\ref{TwoVsThree}(e),(f) demonstrate the resonance shift as a function of the ratio $\Omega_{3}/\Omega_{1}$ for $\Omega_{2}/(2\pi)=1000$~MHz, $\Delta_{1}/(2\pi)=10$~MHz, $\Delta_{2}/(2\pi)=-200$~MHz, which is close to the conditions of our experiment. The calculation is made for the same 200~ns duration of laser $\pi$-pulse, neglecting finite lifetimes of Rydberg and intermediate excited states. The direct numerical simulation of the light shift in a four-level model as a function of $\Omega_{3}/\Omega_{1}$, shown in Fig.~\ref{TwoVsThree}(f), agrees  with the Eq.~\eqref{eq:shifts}. Even in non-perfect conditions of non-zero detunings from the intermediate states, light shift in a three-photon configuration is strongly suppressed.

When the electric field is aligned with the vector of light polarization, only one Stark component of the Rydberg state is excited.
When the electric field is orthogonal to the vector of light polarization, $\sigma$ transitions are allowed. The comparison of Rabi frequencies for different polarizations are summarized in Table~\ref{tab:rabi_frequencies}~\cite{sobelman1992atomic}.

\begin{table}[htbp]
\centering
\caption{Comparison of Rabi frequencies  for transitions $j=1/2 \to j=3/2$ for $\sigma$ and $\pi$ polarizations}
\label{tab:rabi_frequencies}
\begin{ruledtabular}
\begin{tabular}{l|l|l}

\textbf{Polarization} &  $|m_j| = 3/2$ &  $|m_j| = 1/2$  \\
\textbf{Parallel} ($\pi$, $\Delta m_j = 0$)       & $0$ (Forbidden) & $\propto \sqrt{\frac{2}{3}}$  \\
\textbf{Orthogonal} ($\sigma^\pm$, $\Delta m_j = \pm 1$) & $\propto \frac{1}{\sqrt{2}}$ & $\propto \frac{1}{\sqrt{6}}$ \\

\end{tabular}
\end{ruledtabular}
\end{table}

The factor of two variation in the Rabi frequency for laser excitation of the $|m_j|=1/2$ component of the Rydberg state will result in the less than 1~MHz light shift, which is beyond the spectral resolution of our experiment. However, for complete suppression of the light shifts zero detunings $\Delta_1=\Delta_2=0$ are required.

\section{Experiment}
\label{sec.Experiment}

\begin{figure}[!t]
\includegraphics[width= \columnwidth]{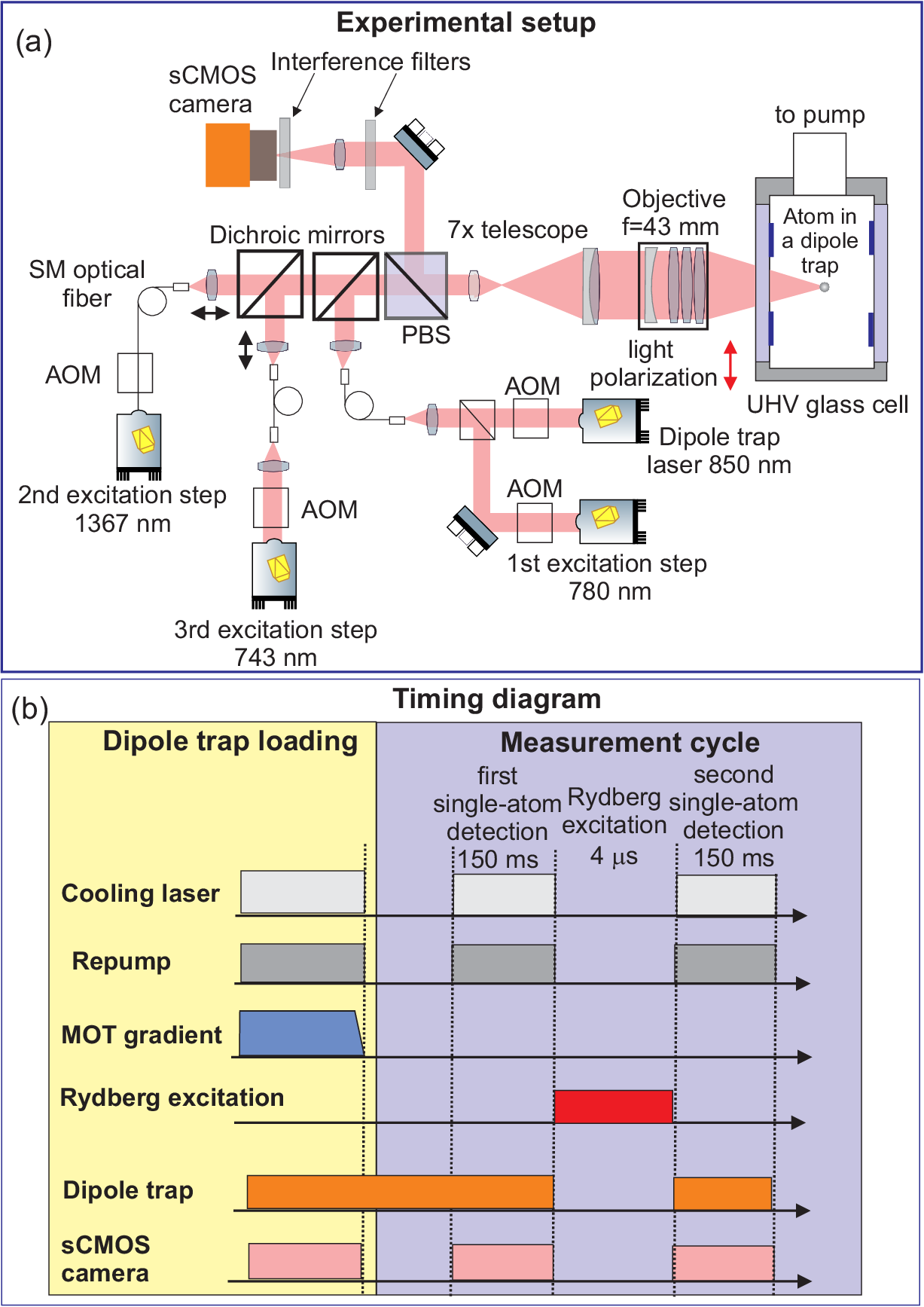}
\caption{(a) Scheme of the experimental setup. Three beams of laser radiation exciting Rydberg atoms are overlapped with the beam of the optical dipole trap and focused into the MOT thorugh the window of the UHV glass cell. A sCMOS camera is used for detection of single atoms using resonance fluorescence. (b) Timing diagram of the experiment. Single Rb atoms are loaded into the optical dipole trap. After detection of single atom, the gradient magnetic field is switched off, and the measurement cycle starts. The single-atom retention is measured after each pulse of Rydberg excitation.}
\label{SetupDetailed}
\end{figure}

The scheme of the experimental setup, described in detail in our previous papers~\cite{Beterov2023,Beterov2024} on three-photon
laser excitation of a single \textsuperscript{87}Rb Rydberg atom, is shown in Fig.~\ref{SetupDetailed}(a).

The magneto-optical trap holds \textsuperscript{87}Rb  atoms at a temperature of 80-100 $\mu$ K. To trap single atoms in an optical dipole trap, we employ 850~nm laser light from an Eagleyard EYP-DFB-0852 DFB master laser, amplified by a Toptica Boosta Pro tapered amplifier to an output power of 1.4~W. An Isomet acousto-optic modulator is used to switch off the trap light during Rydberg excitation. After exiting the polarization-maintaining optical fiber, the dipole trap beam is collimated, reflected from a dichroic mirror, passed through a polarizing beam splitter, and focused into the cloud of cold rubidium atoms by an objective lens with a focal length of 43~mm and a numerical aperture of NA = 0.39, with correction to the 5~mm thick window of a glass cell. A beam expanding telescope consisting of two lenses (f = 75~mm and f =500~mm) is placed in front of the objective. This arrangement focuses the dipole trap radiation to a beam waist radius of approximately 3~$\mu$m.

To detect the trapped atoms, we collect resonance fluorescence at 780~nm induced by the cooling lasers. The emitted photons are collected by the same objective, partially reflected by a polarizing beam splitter, and focused by a lens with a focal length of 50~mm onto a Tucsen Dhyana 400D sCMOS camera. Interference filters are placed in front of the camera to transmit only radiation at 780~nm.

We use three different lasers for three-photon Rydberg excitation. All three lasers are locked to the transmission peak of a highly stable Fabry–Perot interferometer (Stable Laser Systems) using the Pound–Drever–Hall technique. An offset lock is applied at the final excitation step to scan the laser frequency across the resonance. 

The first step of the Rydberg excitation employs a Toptica DL Pro external-cavity diode laser together with a Toptica Boosta Pro tapered amplifier. The radiation from this laser is used both for laser cooling and for Rydberg excitation. Although the detuning of the first-step laser can be controlled independently, in the experiment it was kept equal to the detuning of the cooling radiation, i.e., red-detuned by $\Delta_1=10$~MHz from the resonance with the $\ket{5P_{3/2}\,F=3}$ state. The first-step excitation beam was combined with the dipole-trap radiation in front of the optical fiber using a dichroic mirror.

The second excitation step uses a Sacher Lasertechnik external-cavity diode laser with a wavelength of 1367~nm. The blue two-photon detuning, $\Delta_2=-200$~MHz, was determined by the position of the transmission peak of a high-finesse cavity.

The third step uses a single-frequency Ti:Sapphire ring-resonator laser from Tekhnoscan, pumped by a CNI 532~nm, 10~W single-frequency laser. Its frequency is locked using sideband components generated by mixing a radio-frequency (RF) signal with an arbitrarily chosen frequency in the range from 10~MHz to 200~MHz and applying it to the input of the electro-optic modulator in the frequency-stabilization system. The RF signal is synthesized by a Rigol DG4202 generator controlled via a LAN interface.

The estimated linewidth of all lasers was below 30~kHz. However, in the experiment we observed significantly broader resonances, with a width of approximately 3~MHz. This broadening can be attributed to the residual magnetic field. The radiation frequency is monitored with a precision wavelength meter, WS-U, from Angstrom.

The laser beams used for the second and third steps of the Rydberg excitation are overlapped with the dipole-trap beam and focused onto the same point, with beam waists below 5~$\mu$m. To achieve this, the angular divergence of each beam is adjusted individually.

The timing diagram of the experiment is shown in Fig.~\ref{SetupDetailed}(b). The experimental setup is controlled by a SpinCore PulseBlaster programmable timer board. ${}^{87}\mathrm{Rb}$ atoms are initially loaded into the MOT for 0.1--2~s, while the optical dipole trap is loaded simultaneously. A Tucsen Dhyana 400D digital sCMOS camera detects the atoms as a sequence of images with an exposure time of 150~ms, until a single atom is loaded into the trap and the first resonance-fluorescence signal from the trapped atom appears.

After the detection of a single atom, the procedure of laser excitation to Rydberg states and optical detection of the Rydberg excitation is initiated. The cooling lasers and the MOT gradient magnetic field are switched off. Then the cooling beams, the repumping laser, and the video camera are switched on for the first fluorescence detection from the trapped atom, in order to confirm that the atom remains confined in the optical dipole trap. After that, the dipole-trap radiation is switched off to eliminate the light shifts associated with this radiation, and all three excitation lasers for Rydberg excitation are switched on. After 4~$\mu$s, the excitation lasers are turned off and the optical dipole trap is switched on again.

The intense radiation of the dipole-trap laser ejects the ${}^{87}\mathrm{Rb}$ atom from the trap if it is in a Rydberg state, whereas the atom is recaptured if it remains in the ground state. The cooling laser beams, the repumping laser, and the video camera are then switched on again. An atom that was not excited to a Rydberg state and remained in the optical dipole trap is detected once more. The measurement cycle is repeated until the atom is finally lost from the trap. In single-atom experiments, the single-atom survival probability (single-atom retention) is measured as a function of the excitation-laser frequency.
During the experiment, a voltage in the range of $\pm 10$~V was applied to two  groups of individually selected electrodes, with four electrodes in each group. For this purpose, we used a Keysight 33600A Series low-noise arbitrary waveform generator.

\section{Experimental results and comparison with theory}
\label{sec.Results}

We recorded the spectra of single-atom Rydberg excitation for three different configurations of voltage applied to electrodes, corresponding to all three orthogonal directions of the electric field. The measured dependences of the position of the center of the resonance on the applied voltage are shown in Fig.~\ref{StarkShifts}. The method of calculation of the electric field inside the cell is explained in Appendix~\ref{sec.Field}.  
\begin{figure}[!t]
\includegraphics[width=\columnwidth]{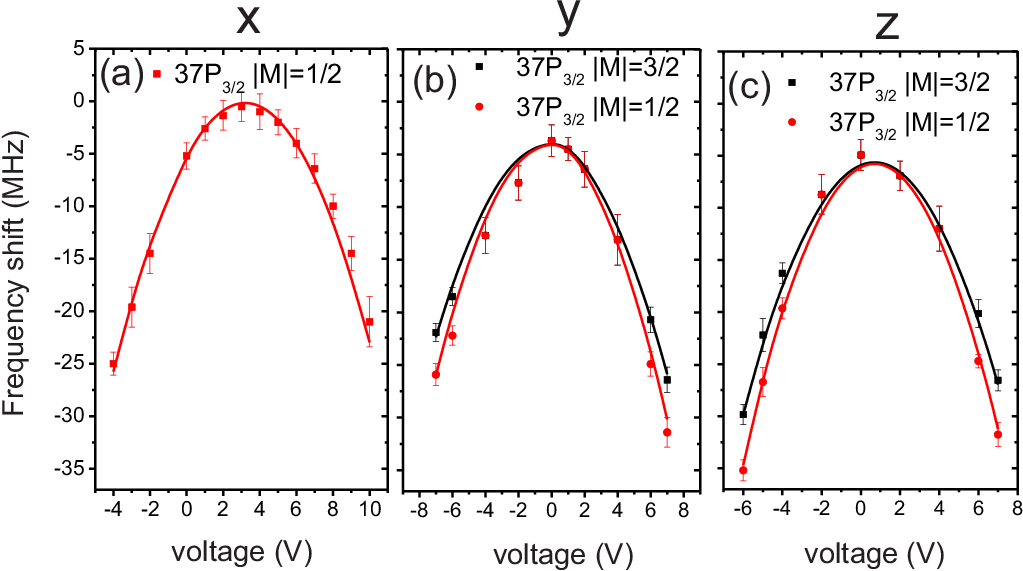}
\caption{ The measured Stark shifts of the resonances when the electric field is pointed to (a) \textit{x} direction, (b) \textit{y} direction, (c) \textit{z} direction.}
\label{StarkShifts}
\end{figure}

When the electric field is directed along the \textit{x} axis, which corresponds to the direction of laser polarization, the positions of the resonances are shifted, as shown in Fig.~\ref{StarkShifts}(a), but no splitting of the resonances is observed.  When the electric field is directed along \textit{y} or \textit{z} axes, which are orthogonal to the direction of laser polarization, the resonances are both shifted and split, as shown in Fig.~\ref{StarkShifts}(b) and Fig.~\ref{StarkShifts}(c). 

In all cases the positions of resonances as functions of applied voltage are approximated by quadratic dependences. These approximations allow for calibration of the electric field in each direction, using the numerically calculated polarizabilities. The position of the minimum of the approximated dependence indicates the amplitude of stray electric field.

From the quadratic approximation of the dependence in Fig.~\ref{StarkShifts}(a) with $\alpha_{P1/2}=19.4$~MHz/(V/cm)\textsuperscript{2} we obtained the following calibration of the electric field related to applied voltage: $E_x \,\mathrm{(V/cm)}=0.16\, U\,\mathrm{(V)}$.  This is consistent with our numeric simulation of the electric field, while the value of the measured electric field being 15\% higher than the calculations. Notably, the minimum of the approximated shift in Fig.~\ref{StarkShifts}(a) is observed when an additional voltage 3.2~V is applied to the electrodes. This corresponds to a stray electric field of around 0.5~V/cm along the axis of the vacuum system, which corresponds to the \textit{x} direction, as shown in Fig.~\ref{Setup}(a).

The approximation of the dependence in \textit{y} direction in Fig.~\ref{StarkShifts}(b) gives similar calibration of the electric field $E_y \,\mathrm{(V/cm)}=0.16 V\,\mathrm{(V)}$  and the smallest value of stray electric field, of approximately 0.05~V/cm.

For \textit{z} direction from the dependence on Fig.~\ref{StarkShifts}(c) we obtained calibration $E_z \,\mathrm{(V/cm)}=0.185\, U\,\mathrm{(V)}$, which gives substantially larger electric field amplitude than the simulation result. In the latter case the discrepancy between the experiment and theory can be attributed to imperfect positioning of the atom in the center between the electrodes. The geometric center of the system corresponds to the minimum of the axial electric field. The smaller, but non-negligible stray electric field of 0.12~V/cm, was also observed in the axial direction along the \textit{z} axis, as shown in Fig.~\ref{StarkShifts}(c).

\begin{figure}[!t]
\includegraphics[width=\columnwidth]{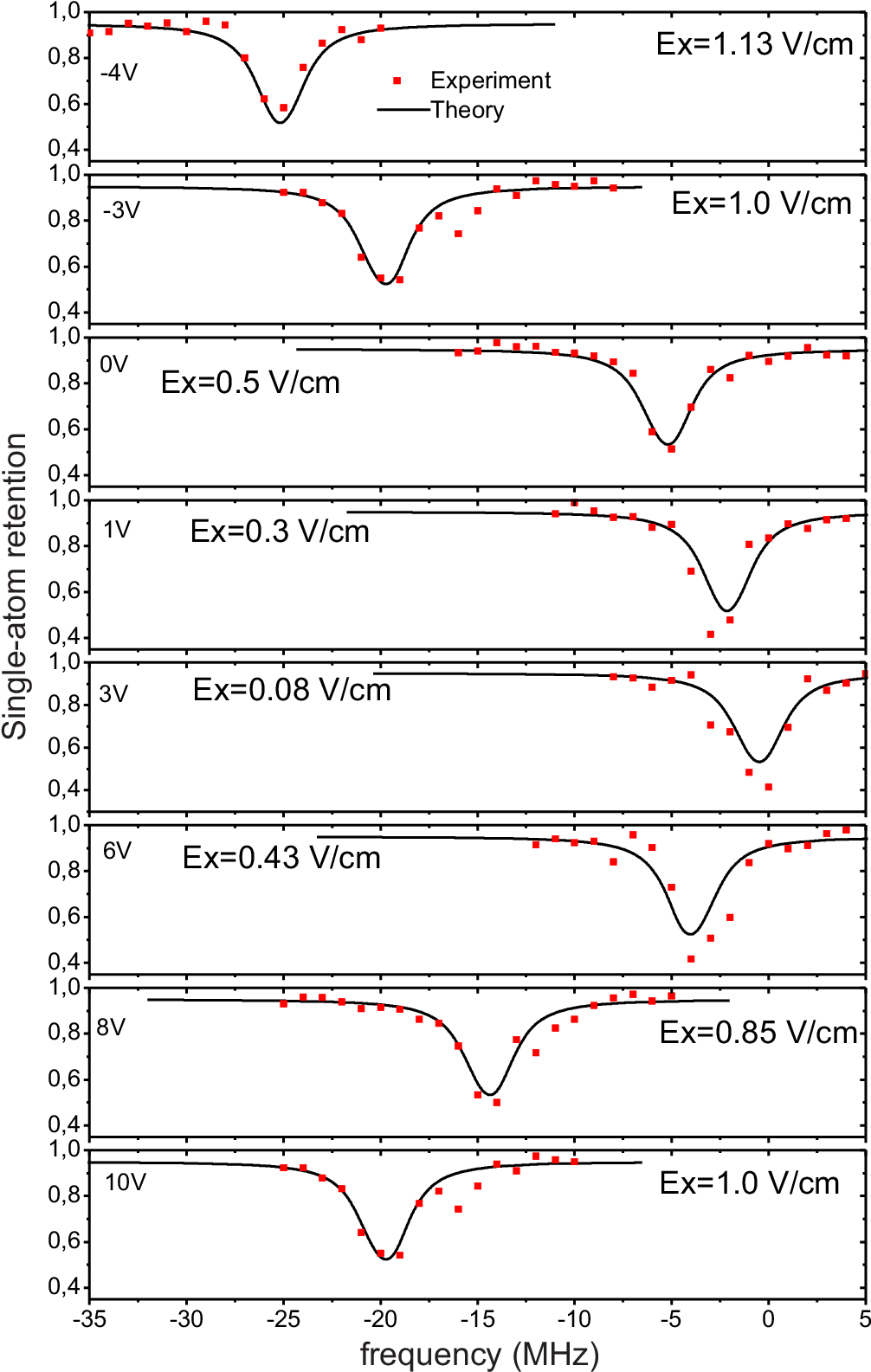}
\caption{ Comparison of experiment and theory for Stark spectra of single-atom Rydberg excitation, when the electric field is scanned across \textit{x} direction. The stray electric fields in the simulation are $E_y=0.05$~V/cm and $E_z=0.12$~V/cm.
}
\label{XSpectra}
\end{figure}

\begin{figure}[!t]
\includegraphics[width=\columnwidth]{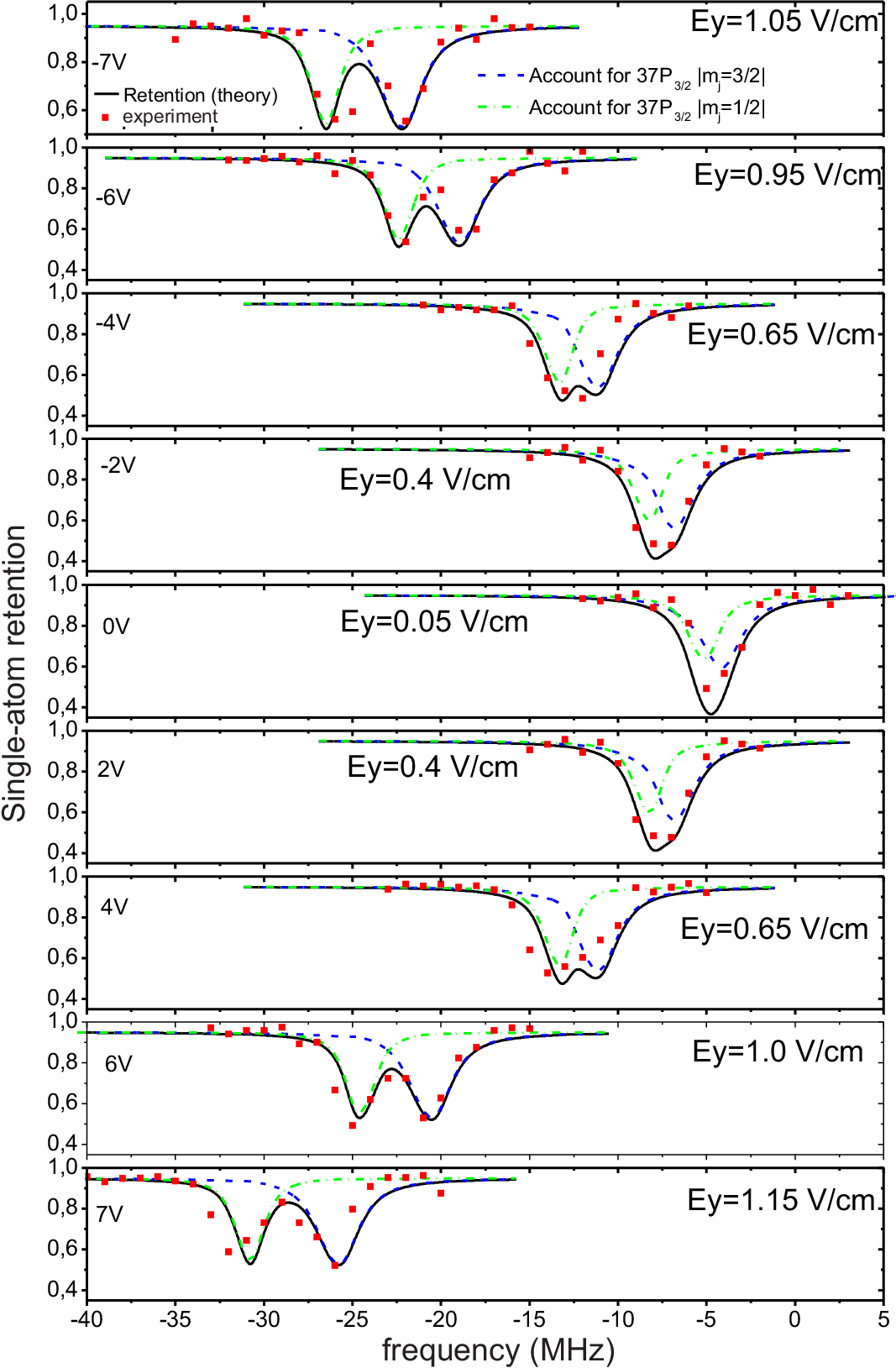}
\caption{ Comparison of experiment and theory for Stark spectra of single-atom Rydberg excitation, when the electric field is scanned across \textit{y} direction.  The stray electric fields in the simulation are $E_x=0.5$~V/cm and $E_z=0.12$~V/cm. Blue (dashed) lines illustrate the account from Rydberg excitation of 37\textit{P}\textsubscript{3/2} state with $|m_j|=3/2$, while green (dash-dotted) lines illustrate the account from Rydberg excitation of 37\textit{P}\textsubscript{3/2} state with $|m_j|=1/2$.
}
\label{YSpectra}
\end{figure}

\begin{figure}[!t]
\includegraphics[width=\columnwidth]{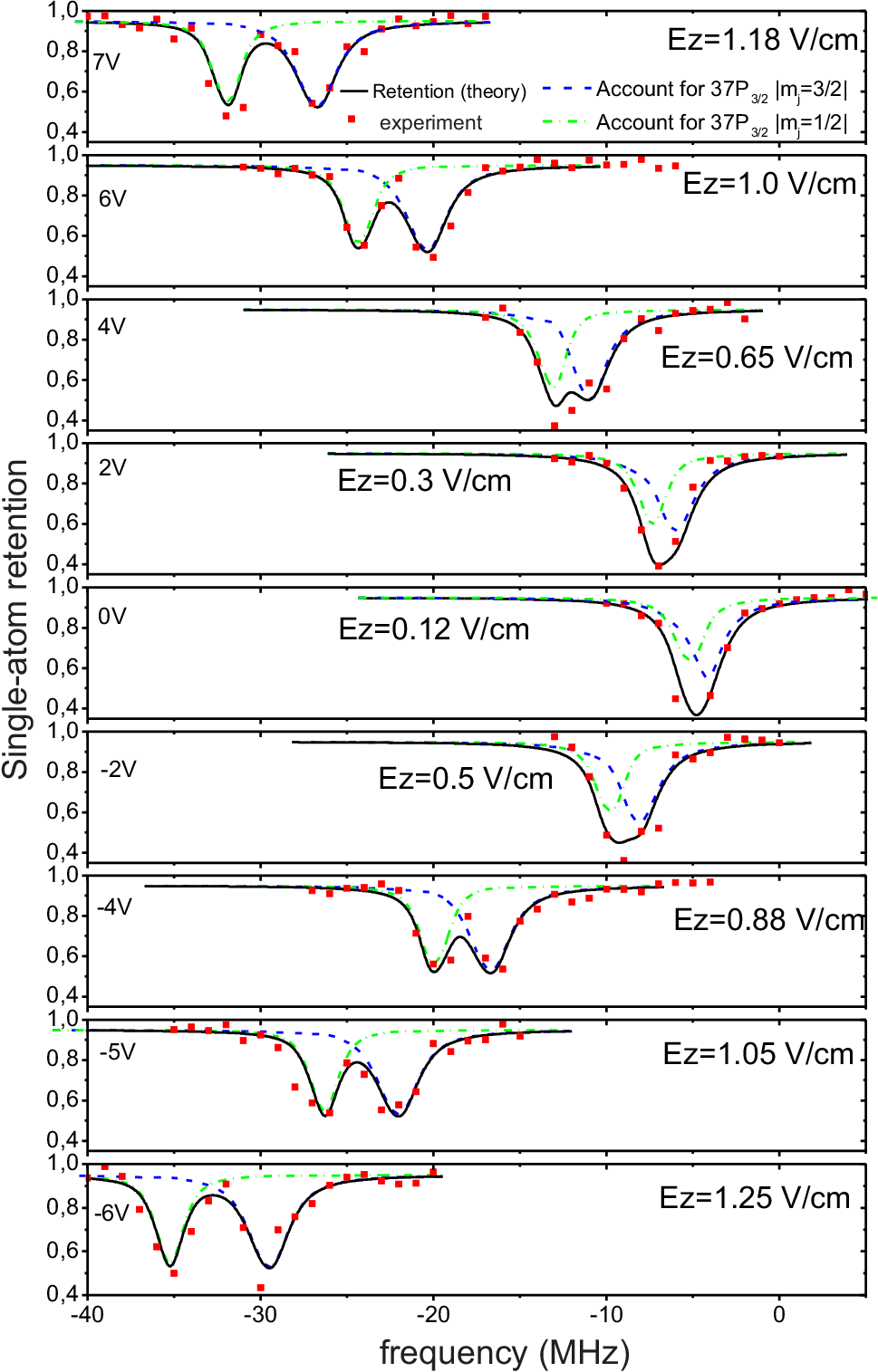}
\caption{ Comparison of experiment and theory of Stark spectra of single-atom Rydberg excitation, when the electric field is scanned across \textit{z} direction.  The stray electric fields in the simulation are $E_x=0.5$~V/cm and $E_y=0.05$~V/cm. Blue (dashed) lines illustrate the account from Rydberg excitation of 37\textit{P}\textsubscript{3/2} state with $|m_j|=3/2$, while green (dash-dotted) lines illustrate the account from Rydberg excitation of 37\textit{P}\textsubscript{3/2} state with $|m_j|=1/2$.
}
\label{ZSpectra}
\end{figure}

To confirm our findings, we numerically calculated the spectra of Rydberg excitation and compared them with the experiment. For our theoretical calculations we estimated the Stark shift of the resonance due to stray electric field using the asymmetry in the dependence of Stark shift on applied voltage from all three records shown in Fig.~\ref{StarkShifts}. This allowed us to calibrate the absolute position of zero Stark shift in Figs.~\ref{StarkShifts}-\ref{ZSpectra}.

We estimated Rabi frequencies $\Omega_{12}/(2\pi)=10$~MHz and $\Omega_{23}/(2\pi)=800$~MHz from the laser power measurements and the  laser beam waists.
For the third step of laser excitation in the case of $\pi$ polarization of the laser beam only the Stark component $\ket{4}=\ket{37P_{3/2}\,|m_j|=1/2}$ is excited, with the  estimated Rabi frequency  $\Omega_{34}/(2\pi)=56$~MHz. 

In the case of $\sigma$ polarization, for the same laser intensities  we obtain Rabi frequencies  $\Omega_{34}/(2\pi)=28$~MHz and $\Omega_{35}/(2\pi)=48$~MHz. Finite laser linewidths and lifting of the degeneracy of atomic energy levels by residual magnetic fields are included in the simulation as pure dephasing processes, with the 300~kHz rate at each transition. Although we used a microwave spectrocopy of the $\ket{5S_{1/2}\,F=1} \to \ket{5S_{1/2}\,F=2} $ transition for partial compensation of residual magnetic fields, they remain present in our experiment.
 
We calculated the single-atom retention taking into account the typical in our experiment approximately -5\% single-atom losses after the first time when the atom was imaged. The simplest case corresponds to scanning the electric field in the \textit{x} direction, when only one of the Stark components $\ket{37P_{3/2}\,|m_j|=1/2}$ is excited. We assumed presence of the non-compensated stray electric fields in \textit{y} and \textit{z} directions. Although these stray fields should result in the excitation of the $\ket{37P_{3/2}\,|m_j|=3/2}$ Stark component, the excitation probability is small enough to be not taken into account. However, in our simulations we took into account the Stark shifts induced by \textit{y} and \textit{z} electric fields, with the amplitudes defined from the approximations in Fig.~\ref{StarkShifts}. Comparison between the theory and experimental records is shown in Fig.~\ref{XSpectra}. In most cases our theoretical model correctly describes both the positions and shapes of the resonances. 

A more interesting case is when the electric field is scanned in orthogonal directions, allowing excitation of both $\ket{37P_{3/2}\,|m_j|=1/2}$ and $\ket{37P_{3/2}\,|m_j|=3/2}$ Stark components. We independently calculated the probability of laser excitation of both components, although they could not be directly distinguished in the experiment, as our atom-loss detection scheme does not have Rydberg state selectivity. For comparison with the experiment we also calculated total population of the Rydberg state. In the simulation we assumed the additional shift of the resonances by the stray electric fields pointed in $x$ and $z$ directions.

Comparison of the theoretical calculations with the experiment is shown in Fig.~\ref{YSpectra}. With the increase of the electric field the resonances at first broaden due to the lifting of the degeneracy of two Stark components of Rydberg state, while the resonances  remain overlapped.

With further increase of the electric field to the values around 1~V/cm the splitting of the resonances is clearly observed. The positions and shapes of the experimentally recorded spectra are well described by our simple theoretical model. Importantly, both shifts and splittings of the resonances are correctly described by the theory, when the numerically calculated polarizabilities are used.

The scanning of the electric field across the \textit{z} axis results in almost similar picture, as illustrated in Fig.~\ref{ZSpectra}.
Despite the asymmetry of the dependence of shifts on the applied voltage, resulting from stray electric field, the positions and shapes of the resonances remain consistent with the theoretical model. This ensures that the shifts of the resonances, when zero voltage is applied to the electrodes, can be attributed to stray DC electric fields. The stray electric fields, observed in our experiments, may arise from the deposition of Rb on the inner surface of the glass cell~\cite{Hattermann2012,Sedlacek2016,Abel2011,Davtyan2018,Chan2014,Mamat2024}. They can be compensated using the electrodes in our cell, thus allowing us to control the electric field of an arbitrary direction to a high precision.

\section{Discussion}
\label{sec.Discussion}

Quantum computing, simulations and electrometry nowadays are the most promising applications of Rydberg atoms. Both tasks require precise control of the electric field in the environment, which includes both suppression of stray electric fields and application of the external DC fields. In the early work~\cite{Frey1993} the residual electric field was suppressed with extremely high precision to the values below $50\,\mu$V/cm using potassium Rydberg states with $n=250$. However, for this purpose the collimated atomic beam  was used, and the interaction region was located inside the large conducting cube with 10~cm side and small openings in its walls. This arrangement reduces the effects of alkali-metal deposition on the inner surfaces and other sources of stray electric fields. However, it limits the optical access, which is required in modern experiments on laser cooling and trapping of single atoms. The opposite situation is observed in modern experiments with atom chips, when the atoms are located at distances of hundred microns from the atom chip surface, and the stray electric fields may reach tens of volts per cm~\cite{Davtyan2018}. 

Many modern experiments on quantum computing and quantum simulations with large-scale atomic arrays are performed in small-size glass cells, allowing optical access for high numeric aperture microscope objectives~\cite{Chiu2025, Manetsch2024, Gyger2024}. In these experiments, no data regarding the presence of stray electric fields were reported. However, we believe that this issue is important for future improvement of  entanglement fidelity between the atoms~\cite{Mamat2024}. From our experimental results, it is clear that stray electric fields in UHV glass cells can be substantially high. For high-fidelity quantum gates, the detuning errors should be kept below few tens of kHz. As an example, Rydberg 70S state has a static polarizability 526~MHz/(V/cm)\textsuperscript{2}~\cite{Vsibalic2017ARC}. This corresponds to stray electric fields of tens of mV/cm.  The methods of suppression of these detrimental electric fields were recently analyzed~\cite{Deng2026}.

Electrodes for precise three-dimensional control of the electric field  in the experiment are common in cold-atom experiments with MOTs in metal vacuum chambers~\cite{Sassmannshausen2013,Ravets2014, Ding2018, Hofmann2014}. They are rarely used in experiments with large-scale atomic arrays~\cite{Pichard2024}. The glass cell with electrodes inside was mentioned in the description of a recent Rb-Cs single-atom experiment~\cite{Miles2026}. We believe that this experimental approach is advantageous for future experiments with large-scale atomic arrays. The size of the UHV cell, used in our experiment, is a compromise between the need for good optical access and the requirement to keep the atoms as far as possible from all surfaces.

In the center of the glass cell the field components are exactly orthogonal by design. However, if the dipole trap is placed out of the center of the cell, the distribution of the electric field components created by different electrode pairs, slightly changes. Radial or axial displacement by 2-3 mm results in approximately 1\% undesirable rotation of the electric field vector, which is hard to observe in our experiment. 

During long-time experiments we observed several jumps in the values of stray electric fields by several MHz, which were restored after several days. We guess that this can be the result of the deposition of Rb atoms on the glass cell walls, when the Rb source is on for a long time. That means that careful control on the positions of the resonances during the experiment is necessary. 

The three-photon scheme of Rydberg excitation of ultracold Rb atoms was used in  our previous experiments with single atoms in an optical dipole trap~\cite{Beterov2023, Beterov2024} and in a MOT~\cite{Entin2013, Ryabtsev2016,Yakshina2018, Tretyakov2022}. In particular, three-photon Rydberg excitation of Rb atoms was demonstrated using only diode lasers in all three excitation steps~\cite{Entin2013}. A similar excitation scheme can also be used for Cs atoms~\cite{Bai2022}. The advantages of three-photon Rydberg excitation were discussed in our recent theoretical work~\cite{Bezuglov2025}. 

\section{Conclusion}
\label{sec.Conclusion}

In summary, we  performed an experiment on Stark spectroscopy of three-photon laser excitation of a single Rb Rydberg atom in an optical dipole trap, scanning the external electric field in all three directions. We used single-atom losses in an optical dipole trap to detect single-atom Rydberg excitation. We have experimentally demonstrated the ability to control the electric field  in an ultrahigh-vacuum glass cell with eight segmented ring electrodes deposited on the inner surface.  By measuring the Stark shifts, induced by the electric field in all three directions, and using numerically calculated polarizabilities of the 37\textit{P}\textsubscript{3/2} Rb Rydberg state, we calibrated the electric field inside the vacuum cell. 

We relied on the theoretically predicted suppression of light shifts for a three-photon scheme of laser excitation, which substantially simplifies the analysis of the experimental records for calibration of the electric field, compared to commonly used two-photon laser excitation schemes. We derived simple analytical expressions describing light shifts in a four-level system with strongly coupled intermediate states. 

For electric field calibration we used the polarizabilities of Rydberg states numerically calculated from the Stark diagrams. The results of calibration  are  consistent with a simple numeric simulation of the electric field generated by eight segmented ring electrodes. A simple five-level theoretical model well describes the recorded spectra of single-atom Rydberg excitation, illustrating Stark shifts, broadening and splitting of the resonances. We found that stray electric fields in experiments with ultrahigh-vacuum glass cells may reach hundred millivolts per cm, which requires careful compensation. 

Overall, our results are important both for future experiments on high-fidelity Rydberg excitation in the controlled environment, and for applications in Rydberg electrometry.

\begin{acknowledgments}
This work was supported by the Russian Science Foundation grant No 23-12-00067-P. 
\end{acknowledgments}

\section*{data availability}
The data that support the findings of this article are openly available \cite{githubStark}.
\newpage
\appendix

\section{Light shifts in a three-photon laser excitation scheme}
\label{sec.Shifts}
In the rotating--wave approximation, with $\hbar=1$ and the ordered basis the Hamiltonian for three-photon Rydberg excitation is written as 
$\{\ket{g},\ket{1},\ket{2},\ket{r}\}$,
\begin{equation}
H=\begin{pmatrix}
0 & \Omega_1/2 & 0 & 0\\[2pt]
\Omega_1/2 & \Delta_1 & \Omega_2/2 & 0\\[2pt]
0 & \Omega_2/2 & \Delta_2 & \Omega_3/2\\[2pt]
0 & 0 & \Omega_3/2 & \delta
\end{pmatrix}.
\label{eq:H}
\end{equation}

We divide the Hilbert space into a ``slow'' subspace $P=\{\ket{g},\ket{r}\}$ and a
``fast'' subspace $Q=\{\ket{1},\ket{2}\}$~\cite{Macri2023}, which leads to
\begin{equation}
H=\begin{pmatrix} H_{PP} & H_{PQ}\\ H_{QP} & H_{QQ}\end{pmatrix},
\end{equation}

\begin{eqnarray}
H_{PP}=\begin{pmatrix}0&0\\0&\delta\end{pmatrix},\qquad & &
H_{PQ}=H_{QP}^{\dagger}=\frac12\begin{pmatrix}\Omega_1&0\\0&\Omega_3\end{pmatrix},\qquad \nonumber \\ 
H_{QQ}=\begin{pmatrix}\Delta_1 & \Omega_2/2\\ \Omega_2/2 & \Delta_2\end{pmatrix}&&.
\end{eqnarray}

The single structural assumption is that the \emph{strong} field $\Omega_2$ lies
\emph{inside} $H_{QQ}$ and is therefore diagonalised exactly, whereas the weak
fields $\Omega_1,\Omega_3$ lie in the off-diagonal block $H_{PQ}$ and are the
perturbation. Ordinary adiabatic elimination is the special case in which
$H_{QQ}$ is additionally assumed diagonal, and it fails when
$\Omega_2^{2}/4\gtrsim\Delta_1\Delta_2$.

We write the stationary problem $H\psi=E\psi$ with $\psi=(\psi_P,\psi_Q)^{T}$:
\begin{align}
H_{PP}\,\psi_P+H_{PQ}\,\psi_Q&=E\,\psi_P, \label{eq:P}\\
H_{QP}\,\psi_P+H_{QQ}\,\psi_Q&=E\,\psi_Q. \label{eq:Q}
\end{align}
Equation~\eqref{eq:Q} is linear in $\psi_Q$ and is solved outright,
\begin{equation}
\psi_Q=\bigl(E-H_{QQ}\bigr)^{-1}H_{QP}\,\psi_P\;\equiv\;G(E)\,H_{QP}\,\psi_P .
\label{eq:slave}
\end{equation}
The fast amplitudes follow the slow ones through a propagator that
already contains $\Omega_2$ to all orders. Substituting \eqref{eq:slave} into
\eqref{eq:P} gives
\begin{eqnarray}
\;\Heff(E)&=&H_{PP}+\Sigma(E),\nonumber  \\
\Sigma(E)&=&H_{PQ}\,\bigl(E-H_{QQ}\bigr)^{-1}H_{QP}\;
\label{eq:Heff}
\end{eqnarray}
which acts on the two-dimensional space $\{\ket{g},\ket{r}\}$. The residual
$E$-dependence survives because no truncation of the $Q$ block was made.
Algebraically, \eqref{eq:Heff} is the \emph{Schur complement} of $H_{QQ}$ in $H$.

With
\begin{eqnarray}
E-H_{QQ}&=&\begin{pmatrix}E-\Delta_1 & -\Omega_2/2\\[2pt] -\Omega_2/2 & E-\Delta_2\end{pmatrix},  \\
\qquad
D(E)&\equiv&\det\bigl(E-H_{QQ}\bigr)=(E-\Delta_1)(E-\Delta_2)-\frac{\Omega_2^{2}}{4}, \nonumber
\end{eqnarray}
the inverse operator is
\begin{equation}
G(E)=\frac{1}{D(E)}
\begin{pmatrix}E-\Delta_2 & \Omega_2/2\\[2pt] \Omega_2/2 & E-\Delta_1\end{pmatrix}.
\label{eq:G}
\end{equation}

The off-diagonal entry $G_{12}=\Omega_2/2D$ exists only because of
$\Omega_2$. It allows a single elimination step to connect $\ket{g}$ to
$\ket{r}$, which is why the three-photon coupling carries $\Omega_2$ in the
numerator instead of requiring two successive eliminations.

Since $H_{PQ}$ scales row $1$ by $\Omega_1/2$ and row $2$ by $\Omega_3/2$, and
$H_{QP}$ does the same to the columns, we obtain
\begin{equation}
\Sigma(E)=\frac{1}{D(E)}
\begin{pmatrix}
\dfrac{\Omega_1^{2}}{4}\,(E-\Delta_2) & \dfrac{\Omega_1\Omega_2\Omega_3}{8}\\[12pt]
\dfrac{\Omega_1\Omega_2\Omega_3}{8} & \dfrac{\Omega_3^{2}}{4}\,(E-\Delta_1)
\end{pmatrix}.
\end{equation}
The effective two-level Hamiltonian for exact three-photon resonance  becomes
\begin{equation}
\;
\Heff(E)=
\begin{pmatrix}
\dfrac{\Omega_1^{2}}{4}\dfrac{E-\Delta_2}{D(E)} &
\dfrac{\Omega_1\Omega_2\Omega_3}{8\,D(E)}\\[14pt]
\dfrac{\Omega_1\Omega_2\Omega_3}{8\,D(E)} &
\delta+\dfrac{\Omega_3^{2}}{4}\dfrac{E-\Delta_1}{D(E)}
\end{pmatrix}.\;
\end{equation}
The diagonal entries are the light shifts of $\ket{g}$ and $\ket{r}$; the
off-diagonal entry is $\Omega_{\mathrm{3ph}}/2$, half the three-photon Rabi
frequency.

Near the three-photon resonance with $\delta\sim0$  we assume $E=0$, which gives
$D\equiv D(0)=\Delta_1\Delta_2-\Omega_2^{2}/4$. Then we obtain light shifts and the three-photon Rabi frequency:
\begin{eqnarray}
S_g&=&-\frac{\Omega_1^{2}\Delta_2}{4\Delta_1\Delta_2-\Omega_2^{2}},\qquad \nonumber \\
S_r&=&-\frac{\Omega_3^{2}\Delta_1}{4\Delta_1\Delta_2-\Omega_2^{2}},\qquad \nonumber \\
\Omega_{\mathrm{3ph}}&=&\frac{\Omega_1\Omega_2\Omega_3}{4\Delta_1\Delta_2-\Omega_2^{2}}.\;
\label{eq:main}
\end{eqnarray}
The three-photon resonance shift is
\begin{equation}
\delta_3^{\mathrm{res}}=S_g-S_r
=\frac{\Omega_3^{2}\Delta_1-\Omega_1^{2}\Delta_2}{4\Delta_1\Delta_2-\Omega_2^{2}}.
\end{equation}
In the weak-coupling limit $\Omega_2^{2}\ll4\Delta_1\Delta_2$ these reduce to
$S_g\to-\Omega_1^{2}/4\Delta_1$ and $S_r\to-\Omega_3^{2}/4\Delta_2$. This is equal to the result of our previous paper~\cite{Entin2013}, with correction to the sign convention.

\newpage
\section{Electric field in a glass cell}
\label{sec.Field}

\subsection*{Method of Moments}
We performed a simple simulation of the electric field created by eight electrodes inside an ultrahigh-vacuum glass cell. 
In a charge-free dielectric region the electrostatic potential $\varphi(\rvec)$ satisfies Laplace's equation, while on the
conductor surfaces it satisfies a Dirichlet condition (each electrode is an equipotential):

\be
\nabla^2 \varphi(\rvec) = 0
\ee 
\noindent  in the region between conductors,
\be
\varphi(\rvec) = V_k
\ee
\noindent for $\rvec \in S_k,\; k=1,\dots,N_e.$ Here $S_k$ is the surface of the $k$-th electrode held at the known potential $V_k$ (in our problem $N_e = 8$).
Instead of solving the differential equation on a volume grid, we use the integral (source) formulation~\cite{Harrington1993}. Every solution of Laplace's equation with the above boundary values can be represented by an unknown surface-charge density $\sigma(\rvec')$ residing on the conductor surfaces $S=\bigcup_k S_k$. The potential produced by that charge is the single-layer potential

\be
\varphi(\rvec)=\frac{1}{4\pi\varepsilon_0}\int_{S}
\frac{\sigma(\rvec')}{\lvert \rvec-\rvec'\rvert}\,dS'\;
\label{eq:single-layer}
\ee

\noindent which automatically satisfies $\nabla^2\varphi=0$ off the surface and $\varphi\to0$ at infinity. The kernel
\be
G(\rvec,\rvec')=\frac{1}{4\pi\varepsilon_0\,\lvert \rvec-\rvec'\rvert}
\ee
\noindent is the free-space Green's function of the Laplace operator,
$-\varepsilon_0\nabla^2 G=\delta(\rvec-\rvec')$.

Enforcing the Dirichlet condition - letting the observation point $\rvec$
approach the surface - turns Eq.~\eqref{eq:single-layer} into a
Fredholm integral equation of the first kind for the unknown $\sigma$:
\be
\frac{1}{4\pi\varepsilon_0}\int_{S}
\frac{\sigma(\rvec')}{\lvert \rvec-\rvec'\rvert}\,dS'
= V_k, \qquad \rvec\in S_k,\; k=1,\dots,N_e .
\label{eq:bie}
\ee
\noindent This is the electrostatic surface-charge integral equation; the
single-layer potential is continuous across $S$, so no jump term appears
(the jump appears only in the normal derivative, i.e.\ in the field).
Equation~\eqref{eq:bie} is the starting point of the Method of Moments (MoM)~\cite{Harrington1993,Gibson2008}. 

For open-boundary electrostatic problems with thin conductor electrodes in free space, the Method of Moments (Boundary Element Method) is the natural choice because it meshes only the 2D conductor surfaces (not the 3D volume), automatically satisfies the far-field boundary condition $\phi(\infty)=0$, and gives the field analytically everywhere from a single surface solve. Finite elements method would require meshing the entire 3D space out to an artificial truncation boundary, yielding orders of magnitude more unknowns for comparable accuracy.

The MoM converts the integral equation Eq.~\eqref{eq:bie} into a linear
algebraic system by three steps~\cite{Harrington1993,Gibson2008}:

\paragraph{Basis-function expansion.}
The surface $S$ is partitioned into $N$ small patches (boundary elements)
$\{S_i\}_{i=1}^{N}$, and the unknown density is expanded in a set of basis
functions $\{f_i\}$:
\begin{equation}
\sigma(\rvec')\approx\sum_{i=1}^{N}\alpha_i\,f_i(\rvec').
\label{eq:expansion}
\end{equation}
With pulse (piecewise-constant) basis functions
\begin{equation}
f_i(\rvec')=
\begin{cases}
1, & \rvec'\in S_i\\[2pt]
0, & \text{otherwise,}
\end{cases}
\end{equation}
each coefficient $\alpha_i=\sigma_i$ is the (constant) charge density on
patch $i$. In our numeric code each patch $dA_i=r_i\,\Delta r\,\Delta\phi$ carries a uniform $\sigma_i$.

\paragraph{Testing (weighting).}
Substituting Eq.~\eqref{eq:expansion} into Eq.~\eqref{eq:bie} leaves a residual;
we force its projection onto a set of testing functions $\{w_j\}$ to
vanish, i.e.\ we take inner products
$\langle w_j,\cdot\rangle=\int_S w_j(\rvec)\,(\cdot)\,dS$. Using
point matching / collocation, $w_j(\rvec)=\delta(\rvec-\rvec_j)$,
the equation is enforced exactly at the patch centroids $\rvec_j$:
\begin{equation}
\sum_{i=1}^{N} P_{ji}\,\sigma_i = V(\rvec_j),\qquad j=1,\dots,N,
\label{eq:linear}
\end{equation}
with the potential-coefficient (moment) matrix
\begin{equation}
P_{ji}=\frac{1}{4\pi\varepsilon_0}\int_{S_i}
\frac{dS'}{\lvert \rvec_j-\rvec'\rvert}.
\label{eq:Pmatrix}
\end{equation}

\paragraph{Matrix solution.}
Equation~\eqref{eq:linear} is the dense linear system
\begin{equation}
\;\mathbf{P}\,\bm{\sigma}=\mathbf{V}\;\qquad\Longrightarrow\qquad
\bm{\sigma}=\mathbf{P}^{-1}\mathbf{V},
\label{eq:solve}
\end{equation}
where $\mathbf{V}=[V_1,\dots,V_N]^{\mathsf T}$ assigns to each patch the
potential of the electrode it belongs to. $\mathbf{P}$ is symmetric
positive-definite (it is the discrete Coulomb operator), so Eq.~\eqref{eq:solve} has a unique solution.

\paragraph{Off-diagonal terms ($j\neq i$).}
For well-separated patches the integrand is smooth and the
centroid (monopole) approximation is used,
\begin{equation}
P_{ji}\approx\frac{A_i}{4\pi\varepsilon_0\,
\lvert \rvec_j-\rvec_i\rvert},
\qquad A_i=\int_{S_i}dS' ,
\label{eq:offdiag}
\end{equation}
i.e.\ patch $i$ acts as a point charge $q_i=\sigma_i A_i$ at its centroid.

\paragraph{Diagonal terms ($j=i$).}
Here $\rvec_j\in S_i$ and the kernel is singular
($1/|\rvec-\rvec'|$), but the singularity is  integrable. It is
evaluated analytically. Approximating patch $i$ by an equal-area disk of
radius $a_i=\sqrt{A_i/\pi}$, the potential at the center of a uniformly
charged disk is finite:
\begin{equation}
P_{ii}=\frac{1}{4\pi\varepsilon_0}\int_{S_i}
\frac{dS'}{\lvert \rvec_i-\rvec'\rvert}
\;\approx\;\frac{1}{4\pi\varepsilon_0}
\int_0^{a_i}\!\!\int_0^{2\pi}\frac{\rho\,d\rho\,d\theta}{\rho}
=\frac{a_i}{2\varepsilon_0}.
\label{eq:self}
\end{equation}
This is used in the numeric code as self-term. (For a square element of side $h$ the exact result
$P_{ii}=\frac{h}{\pi\varepsilon_0}\ln(1+\sqrt2)$ may be used instead~\cite{Gibson2008,Rao1984}).
Analytic treatment of the singular self- and near-terms is essential for accuracy~\cite{Rao1984}.

Once $\bm{\sigma}$ is known, the potential and field at  any point
$\rvec$ follow from superposition of the patch charges $q_i=\sigma_i A_i$:
\begin{equation}
\varphi(\rvec)=\frac{1}{4\pi\varepsilon_0}\sum_{i=1}^{N}
\frac{q_i}{\lvert \rvec-\rvec_i\rvert},
\label{eq:phi-post}
\end{equation}
\begin{equation}
\;
\Evec(\rvec)=-\nabla\varphi(\rvec)
=\frac{1}{4\pi\varepsilon_0}\sum_{i=1}^{N}
q_i\,\frac{\rvec-\rvec_i}{\lvert \rvec-\rvec_i\rvert^{3}}.
\label{eq:E-post}
\end{equation}
 Other quantities follow directly: the electrode charge
$Q_k=\sum_{i\in S_k}q_i$, the capacitance/Maxwell coefficients
$Q_k=\sum_l C_{kl}V_l$ (with $\mathbf{C}$ obtainable column-by-column from
$\mathbf{P}^{-1}$), and the electrostatic energy
$W=\tfrac12\sum_k Q_kV_k=\tfrac12\,\mathbf{V}^{\mathsf T}\mathbf{P}^{-1}
\mathbf{V}$.

\subsection*{Simulation algorithm}
In our Python code the MoM solver for the 8-electrode system is implemented in four main functions that execute sequentially. These functions are the following:

\textit{I. Geometry Construction:}

Discretizing the two annular rings into $N$ surface patches.

\textbf{Algorithm:}
\begin{enumerate}[leftmargin=2em]
\item Define physical geometry of two coaxial rings. Each ring is divided into 4 quadrants with specified voltages.
\item Create polar mesh: $N_r=8$ radial $\times$ $N_\phi=48$ angular
      subdivisions
\item For each patch $(i_r, i_\phi)$ on each ring:
  \begin{itemize}
  \item Compute centroid:
        $\mathbf{r}_i = (r_c\cos\phi_c,\, r_c\sin\phi_c,\, z)$
  \item Compute area: $A_i = r_c \,\Delta r \,\Delta\phi$
  \item Assign voltage $V_i$ based on which quadrant the patch belongs to
  \end{itemize}
\end{enumerate}

Our code uses $N = 2 \times 4 \times 8 \times 48 = 768$ patches.

\textit{II. Matrix Assembly and Solution:}

Solving the linear system
$\mathbf{P}\,\boldsymbol{\sigma} = \mathbf{V}$ for surface charge
densities.

\textbf{Algorithm:}
\begin{enumerate}[leftmargin=2em]
\item Initialize $N \times N$ dense matrix $\mathbf{P}$
\item For each patch $j = 1, \dots, N$:
  \begin{itemize}
  \item \textbf{Off-diagonal elements} ($i \neq j$):
  \[
  P_{ji} = \frac{A_i}{4\pi\varepsilon_0\,|\mathbf{r}_j - \mathbf{r}_i|}
  \]
    (monopole approximation, vectorized over all $i$ simultaneously)
  \item \textbf{Diagonal element} ($i = j$):
  \[
  P_{jj} = \frac{a_j}{2\varepsilon_0}, \quad a_j = \sqrt{A_j/\pi}
  \]
  (analytic self-potential of equivalent disk)
  \end{itemize}
\item Solve: $\boldsymbol{\sigma} = \mathbf{P}^{-1}\mathbf{V}$ via
      dense LU factorization
\end{enumerate}

\textit{III. Potential Evaluation:}

Computing potential at arbitrary points via Coulomb
superposition.

\textbf{Algorithm:}
\begin{enumerate}[leftmargin=2em]
\item Convert surface densities to point charges: $q_i = \sigma_i A_i$
\item For each evaluation point $\mathbf{r}$:
\[
\varphi(\mathbf{r}) = \frac{1}{4\pi\varepsilon_0}
\sum_{i=1}^{N} \frac{q_i}{|\mathbf{r} - \mathbf{r}_i|}
\]
(vectorized sum over all 768 patches)
\end{enumerate}

\textit{IV. Field Evaluation:}

Computing electric field vector at arbitrary points.

\textbf{Algorithm:}
\begin{enumerate}[leftmargin=2em]
\item For each evaluation point $\mathbf{r}$:
\[
\mathbf{E}(\mathbf{r}) = \frac{1}{4\pi\varepsilon_0}
\sum_{i=1}^{N} q_i\,
\frac{\mathbf{r} - \mathbf{r}_i}{|\mathbf{r} - \mathbf{r}_i|^3}
\]
This is a vectorized 3-component sum which allows calculation of the full electric field distribution.
\end{enumerate}

\textit{Verification}

Re-evaluating potential at patch centroids:
\[
\hat{\mathbf{V}} = \mathbf{P}\,\boldsymbol{\sigma}
\]
should recover the prescribed voltages to $\sim 10^{-4}$ relative error
(limited by monopole vs.\ disk self-term mismatch).

\newpage
\subsection*{Results of simulation}
\begin{figure}[!t]
\includegraphics[width=\columnwidth]{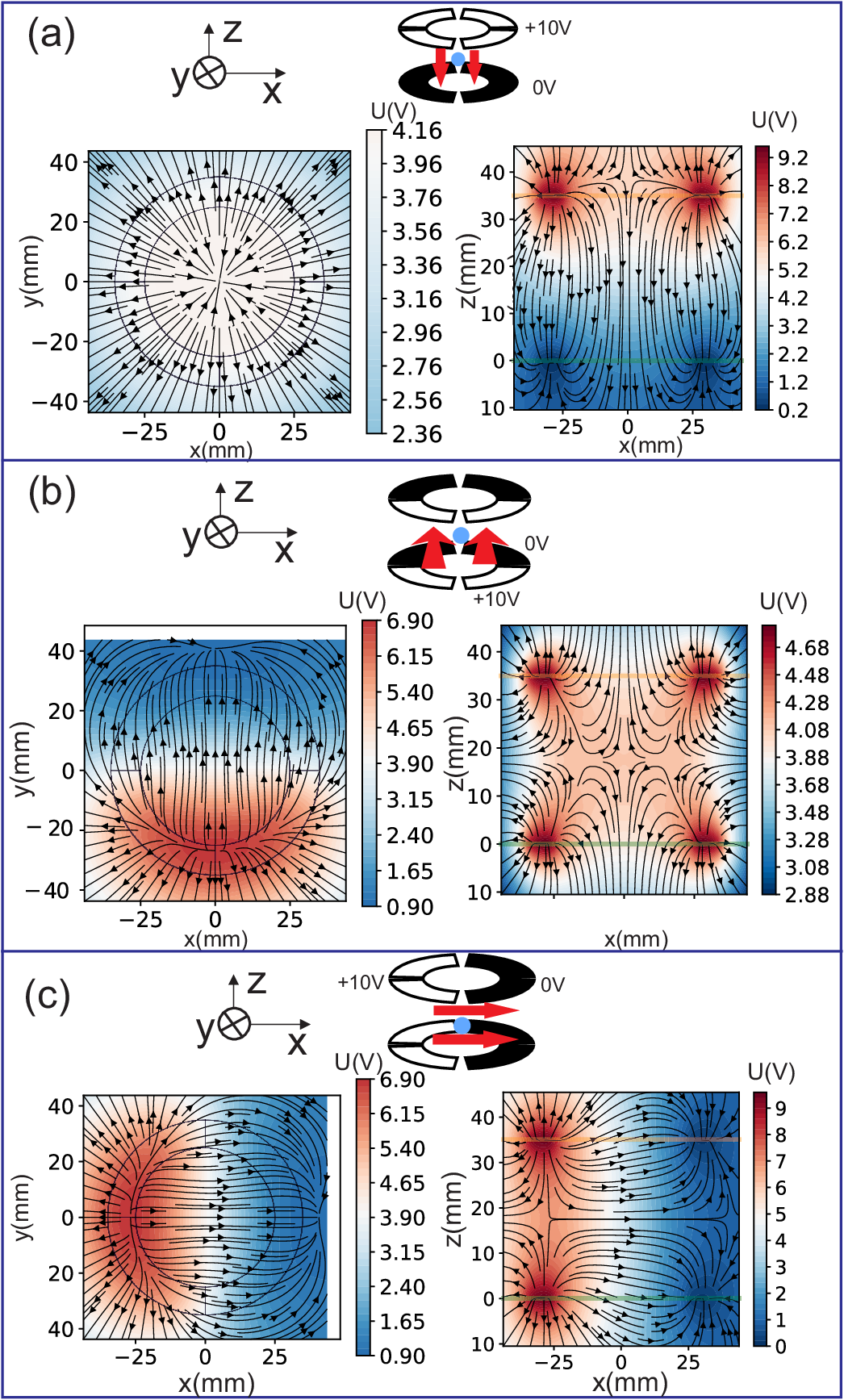}
\caption{The calculated distribution of electric potential and directions of the electric field in the \textit{xy} and \textit{xz} planes passing through the geometric center of the system. (a) The electric field is pointed to the \textit{z} direction, when voltage $U=+10$~V is applied to one of the rings, and the second ring is grounded. (b) The electric field is pointed to the \textit{y} direction, when voltage $U=+10$~V is applied to the pair of neighboring segments of each ring, and the remaining segments are grounded. (c) The electric field is pointed to the \textit{x} direction, when voltage $U=+10$~V is applied to the other pair of neighboring segments of each ring, and the remaining segments are grounded.
}
\label{ElectricField}
\end{figure}

The calculated distribution of electric potential and directions of the electric field in the \textit{xy} and \textit{xz} planes passing through the geometric center of the system are shown in Fig.~\ref{ElectricField} for three different configurations of voltage applied to the electrodes. The distance between the rings is 35~mm, the inner and outer radii of the rings are 25~mm and 35~mm, respectively.

In the configuration of Fig.~\ref{ElectricField}(a) all segments of the first ring are grounded and voltage is applied to all segments of the second ring. In this case the electric field should be directed along the axis of two coaxial rings. 

As we see from the distributions in Fig.~\ref{ElectricField}(a), in the mid-plane \textit{xy} the electrostatic potential is nearly uniform. From the distribution in the \textit{xz} plane it s clear that the electric field is directed along the \textit{z} axis. For the +10~V potential difference between the electrodes the value of the calculated electric field in the geometric center of the system is $E_z=1.39$~V/cm. 

In the configuration of Fig.~\ref{ElectricField}(b) two neighboring segments of each ring are grounded, and voltage +10~V is applied to other two segments of each ring. In this case the electric field is directed along the \textit{y} axis, as can be seen from the  distributions of the potential and electric field. The calculated value of the electric field in the geometric center is $E_y=1.43$~V/cm, which is close to the previously obtained value of the axial field, when the same voltage is applied. The field can be also directed across the \textit{x} axis, as shown in Fig.~\ref{ElectricField}(c). Due to the symmetry, electric field in \textit{x} direction has the same value  $E_x=1.43$~V/cm. 

Our simulation confirms that this geometry of electrodes allows pointing of the electric field in any direction. For the selected geometry of the electrodes (radii and distance) we obtain almost the same ratio between the applied voltage and electric field both in axial and radial directions. We confirmed that these simple calculations are in general consistent with the results of more complex simulations using finite element analysis with COMSOL\textsuperscript{TM} solver.

\bibliographystyle{apsrev4-2}
%

\end{document}